\documentclass[12pt]{article}  
\usepackage{a4,epsfig,axodraw,graphics}
\begin{document}




\begin{titlepage}

\pagenumbering{arabic}
\vspace*{-1.5cm}
\begin{tabular*}{15.cm}{lc@{\extracolsep{\fill}}r}
{\bf DELPHI Collaboration} & 
\hspace*{1.3cm} \epsfig{figure=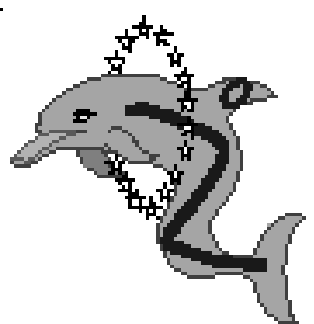,width=1.2cm,height=1.2cm}
&
DELPHI 2003-062 TALK 35
\\
& &
25 June, 2003
\\
&&\\ \hline
\end{tabular*}
\vspace*{1.cm}
\begin{center}
\Large 
{\bf \boldmath
Four-fermion production and limits on anomalous couplings at LEP-2
} \\
\vspace*{1.5cm}
\normalsize { 
   {\bf P. Bambade}\\
   {\footnotesize LAL, Orsay}
}
\vspace*{1.0cm}
\end{center}
\vspace{\fill}
\begin{abstract}
\noindent
Electroweak processes with four fermions in the final state were measured 
extensively at LEP-2 in e+e- collisions with centre-of-mass energies up to 
209~GeV. By combining the results obtained by Aleph, Delphi, L3 and Opal, the 
predictions from the Standard Model were probed at the level of their 
present accuracy.  An important outcome was the possibility to measure 
the predicted charged triple gauge boson self-couplings with a precision of a 
few percent. Moreover studies of four-fermion processes allowed new physics 
to be searched for by looking for anomalous gauge boson self-couplings. 
The measurements reported are also relevant to direct searches for new particles 
at LEP, as four-fermion processes contribute an important background in several 
cases. The latest results from the LEP experiments - some of which are now 
available in finalized form - and their combination are reviewed and discussed.

\end{abstract}
\vspace{\fill}
\vspace*{1.0cm}
\begin{center}
Talk given at Les Rencontres de Physique de la Vall{\' e}e d'Aoste,
la Thuile, Aosta Valley (Italy), March 9-15, 2003.
\end{center}
\vspace{\fill}
\end{titlepage}

\section{Introduction}
During the second part of the LEP program in 1995-2000, the centre-of-mass 
energy of the $\rm e^+e^-$ collisions was raised above the Z boson mass, 
reaching up to 209 GeV in the last year of operation. Integrated 
luminosities of about 700 $\rm pb^{-1}$ were collected by each of the four LEP 
collaborations ALEPH, DELPHI, L3 and OPAL, mostly above the W and 
Z-pair production thresholds. This allowed a comprehensive measurement 
program of boson pair production in $\rm e^+e^-$ collisions and more generally
of all possible four-fermion 
($4f$) final states. Two of the principal aims of LEP,
directly validating the non-Abelian gauge structure of the standard 
model (SM) and searching for anomalous gauge boson self-couplings to 
reveal new physics beyond the SM (NP), could be pursued.

Most of the measured $4f$ final states result from processes 
seen at LEP for the first time.
Initially some of the theoretical predictions and modelling were crude.
Improvements were needed to match the experimental accuracy and were 
challenging, especially in the case of W-pair production\cite{lepmc}. 
The LEP measurements stimulated important work to compute 
radiative corrections and to devise appropriate ways to choose the 
scales of coupling constants when several were involved. 
The results were also included in practical event generators,
where numerical divergences resulting from the small electron mass 
and the handling of the full set of graphs both had to be dealt with
efficiently.

The experience gained at LEP in measuring $4f$ final states and the progress 
in the theoretical description and modelling were also important in
the context of many new particle searches. In several cases the experimental 
signatures are very similar to those of the signal searched for, 
thus providing a useful environment to test the 
corresponding analysis techniques\footnote{A good example is 
the Z-pair production process, for which event rates and 
topologies were almost identical to those in the Higgs boson search.}.
In the longer term, the improvements 
in the calculations will also benefit similar work at a future $\rm e^+e^-$
linear collider, where very high precision will be needed.

This report has two sections. Firstly, the latest 
measurements of $4f$ final states arising via the production
processes $\rm e^+e^- \rightarrow WW$, ZZ, 
$\rm Z\gamma^*$, $\rm eeZ/\gamma^*$ and $\rm e\nu W$ are described. 
Secondly, interpretations of these measurements
in terms of constraints on gauge boson self-couplings are presented,
covering both charged and neutral triple gauge couplings (TGC)
and the more recently studied quartic gauge couplings (QGC).
For each channel, salient experimental, technical or theoretical features 
are highlighted and results from combining the LEP data-sets are presented
when available. The main achievements are assessed after each section, comparing
cross-section measurements to present theoretical predictions, and constraints 
on gauge boson self-couplings to estimates of effects in scenarios for NP,
and to sensitivities expected in the future.
A general conclusion is given with prospects to finalize and publish all results.

\vspace{\fill}

\section{Four-fermion processes}
\subsection{Signal definition in the simulation}
\begin{figure}[!h]
  \vspace{12.0cm}
  \includegraphics{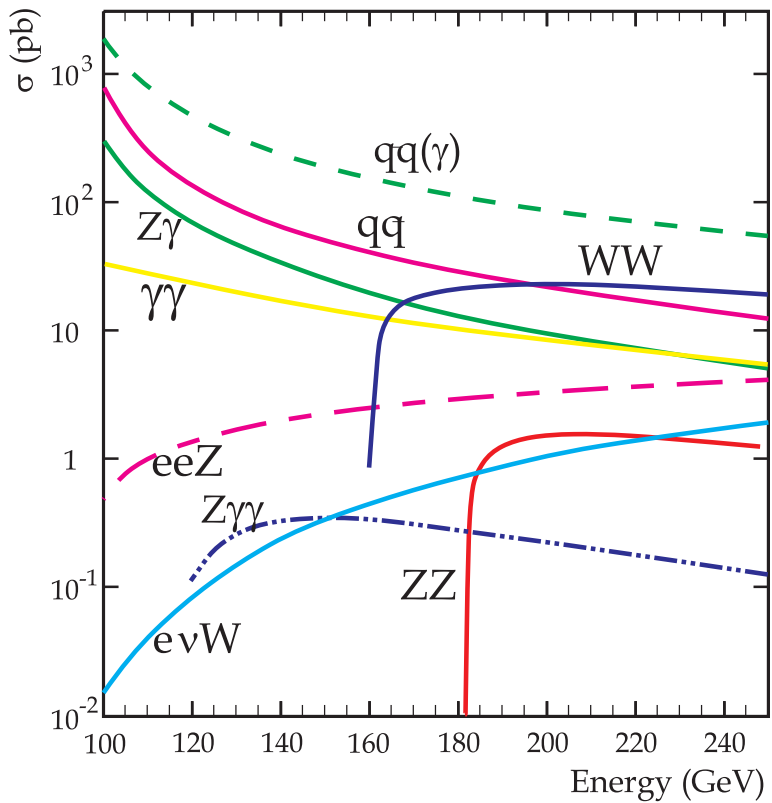}
  \caption{\it
    Cross-sections of the main two and four-fermion processes at LEP-2.
    \label{cross} }
\end{figure}

The expected cross-sections of the production processes 
$\rm e^+e^- \rightarrow WW$, ZZ, $\rm Z\gamma^*$, $\rm eeZ/\gamma^*$ and 
$\rm e\nu W$ are shown for illustration in fig.\ref{cross}. 
When combining the measurements of different sub-channels or experiments,
it is important to establish common conventions. Two ways of defining a
signal to be measured were used, either as the total cross-section in 
a kinematic region where it is dominant or as that corresponding 
exclusively to the relevant Feynman graphs.
In both cases a sub-set of events was pre-selected in the
simulation with high purity and efficiency in terms of 
relative contribution of the studied process. The selected rates 
were then scaled into those of the studied process, taking into
account interference effects for identical final states. 
In the first method cuts were chosen in view of theory
uncertainties affecting both the studied and other contributing 
processes in the different regions. 

\subsection{WW production}
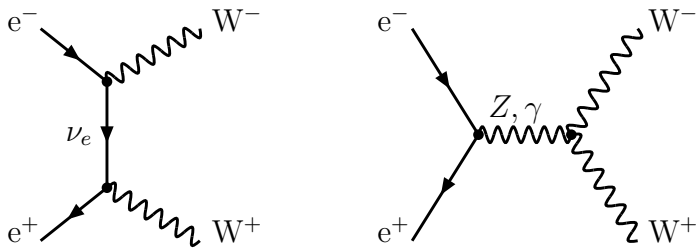
\begin{figure}[!h]
  \begin{center}
     \SetWidth{1.1}
     \begin{picture}(400,100)(0,100)
     \ArrowLine(70,180)(95,160)
     \ArrowLine(95,160)(95,120)
     \ArrowLine(95,120)(70,100)
     \Photon(95,160)(130,180){3}{6}
     \Photon(95,120)(130,100){3}{6}
     \Vertex(95,160){2}
     \Vertex(95,120){2}
     \Text(70,100)[rb]{e$^{+}$}
     \Text(70,180)[rb]{e$^{-}$}
     \Text(135,180)[lb]{W$^{-}$}
     \Text(135,100)[lb]{W$^{+}$}
     \Text(90,140)[r]{$\nu_e$}


     \ArrowLine(210,180)(235,140)
     \ArrowLine(235,140)(210,100)
     \Photon(235,140)(270,140){3}{6}
     \Photon(270,140)(295,180){3}{6}
     \Photon(270,140)(295,100){3}{6}
     \Vertex(235,140){2}
     \Vertex(270,140){2}
     \Text(210,100)[rb]{e$^{+}$}
     \Text(210,180)[rb]{e$^{-}$}
     \Text(300,180)[lb]{W$^{-}$}
     \Text(300,100)[lb]{W$^{+}$}
     \Text(260,150)[r]{$Z,\gamma$}

  \end{picture}
  \vspace{1em}
  \caption{\it Feynman graphs for on-shell WW production.}
  \label{cc03graph}
  \end{center}
  \end{figure}

\begin{figure}[!h]
  \vspace{9.5cm}
  \includegraphics{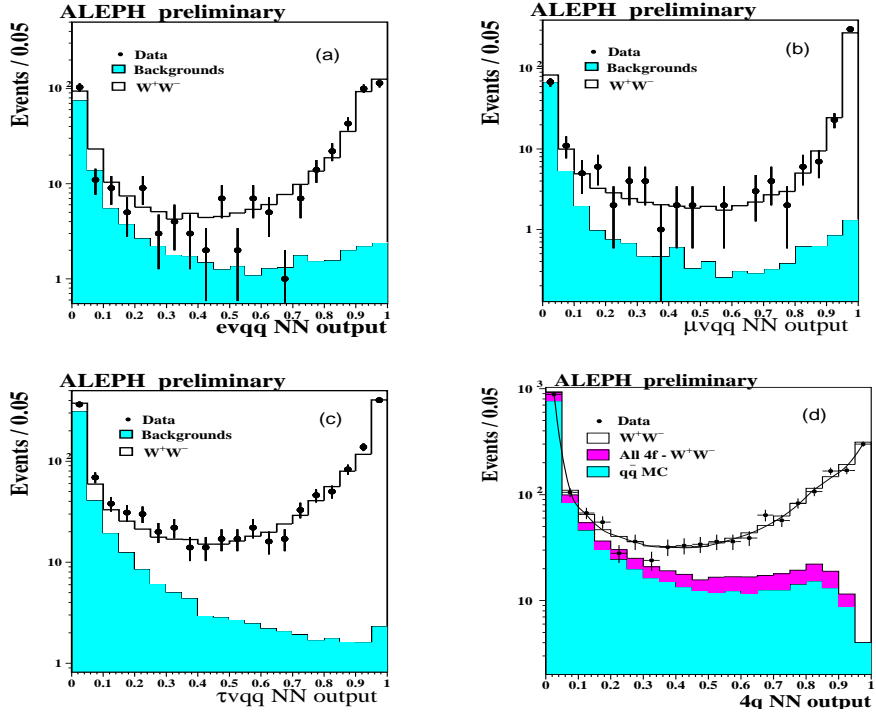}
  \caption{\it
    Outputs from the neural networks used to isolate the W-boson pair
    signal in the fully hadronic and in the three semi-leptonic sub-channels.
    \label{alephww} }
\end{figure}

W-boson pairs are produced via the doubly-resonant 
graphs shown in fig.\ref{cc03graph} at tree level.  
Measurements of cross-sections and decay branching ratios, 
using a total of about 40000 events selected by the four LEP collaborations 
in the data collected in 1996-2000, are reported in\cite{confww}. 
All decay topologies $\rm q \bar{q} q \bar{q}$, $\rm l \nu q \bar{q}$ and 
$\rm l \nu l \bar{\nu}$ were analysed. 
The separation achieved by ALEPH 
between signal and background is illustrated in fig.\ref{alephww}.  

\begin{figure}[!h]
  \vspace{9.8cm}
  \includegraphics{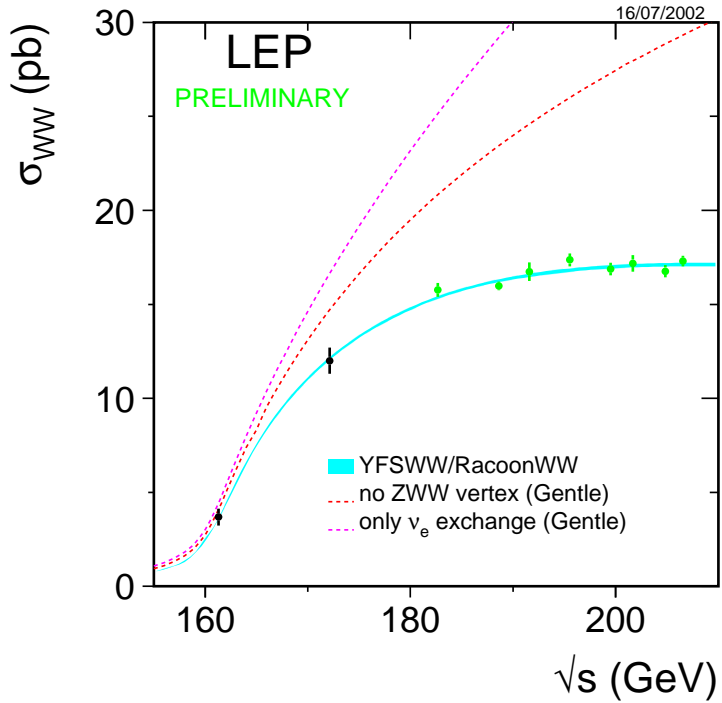}
  \caption{\it
    Combined WW cross-section measurements
    compared with the SM predictions from YFSWW and RacoonWW. 
    The dashed curves correspond 
    to removing one or both of the graphs with triple gauge boson 
    couplings in fig.\ref{cc03graph}.
    \label{wwcross} }
\end{figure}

The cross-sections obtained by combining all LEP results are shown 
in fig.\ref{wwcross} as a function of the centre-of-mass energy, with a 
comparison to theoretical expectations\cite{lep4f}.
The results clearly favour the SM prediction where the graphs in 
fig.\ref{cc03graph} involving TGCs are included, 
hence providing strong evidence for the non-Abelian gauge group structure 
of the theory. The highest precision was achieved by combining the results from
all energies normalised to respective expectations. 
The best agreement was
obtained using the new YFSWW\cite{yfsww} and RacoonWW\cite{racoon} predictions. 
In these two calculations all $O(\alpha)$ electroweak (EW) radiative corrections 
relevant to the graphs in fig.\ref{cc03graph} were included
through expansions about the W-pole 
(in the so-called leading and double pole approximations, respectively). 
They were introduced into ``complete'' $4f$ generators
such as KoralW\cite{kandy} and Wphact\cite{wphact} by reweighting 
the relevant matrix elements, to achieve a consistent 
description of all processes in the full phase space\footnote{The 
matching to the two-photon generation was also studied in Wphact.}.
The result obtained using YFSWW and combining the data from all energies was
\begin{equation}
  {\sigma_{\rm WW} \over \sigma_{\rm YFSWW}} = 0.997 \pm 0.007(stat) \pm 0.009(syst).
  \label{avww} 
\end{equation}
Results with RacoonWW differed by only 0.2\%, well within the estimated theory
uncertainty of 0.5\%. On the contrary a set of previously used predictions 
which did not include all EW radiative corrections were too high by 2\%,
illustrating the sensitivity to loop effects achieved in these results\cite{lepmc}.

The main source of uncertainty was from 
limited precision in the modelling of QCD fragmentation and 
hadronization, which could bias selection 
efficiencies in the semi-leptonic and fully hadronic sub-channels and in the latter, 
background levels from $\rm q \bar{q}(\gamma)$ final states with four or
more jets.
Such effects being correlated among experiments and energies and
of similar magnitude as the statistical errors, a
careful treatment was important in the combination\cite{lep4f}. 
All systematic errors were grouped in four classes 
(100\% correlated or uncorrelated among experiments or energies) 
and the full covariance matrix for the measurements at each experiment 
and energy was built for the $\chi^2$-minimisation.
This procedure was applied to all LEP cross-section combinations 
described in this report.
For the result quoted in
eq.\ref{avww} it gave $\chi^2$ = 35.4 for 31 degrees of freedom.

\subsection{ZZ production}

\begin{figure}[!h]
  \vspace{10.5cm}
  \includegraphics{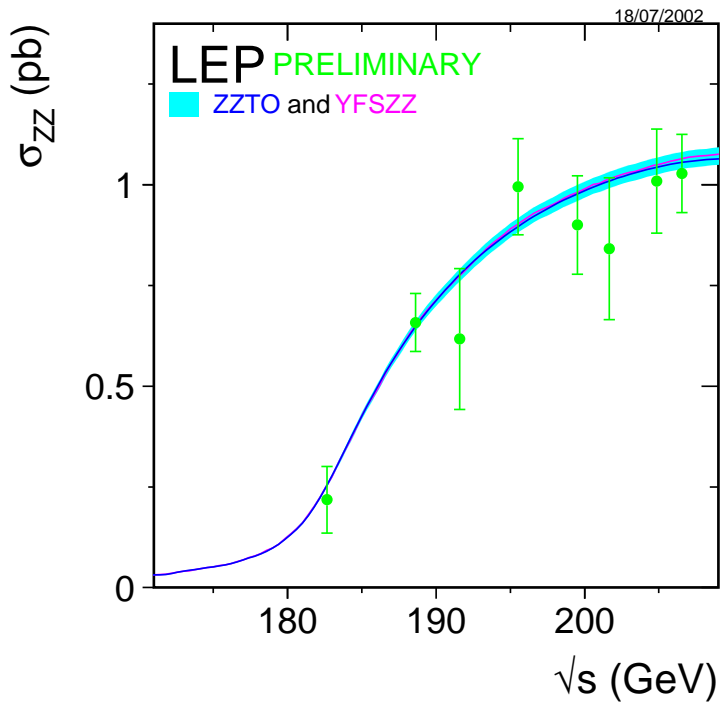}
  \caption{\it
    Combined ZZ cross-section measurements compared with the SM predictions 
    from ZZTO and YFSZZ. The band shows the theory uncertainty.
    \label{zzcross} }
\end{figure}

Z-boson pair production occurs at tree level in the SM only via graphs with 
a $t$-channel electron (similar to that with a $\nu$ in fig.\ref{cc03graph}).
The cross-section was measured by the four LEP collaborations 
with data collected in 1997-2000 using all visible decays
$\rm q \bar{q} q \bar{q}$, $\rm \nu \bar{\nu} q \bar{q}$, 
$\rm l l q \bar{q}$, $\rm l l l l$ and $\rm \nu \bar{\nu} l l$\cite{confzz}. 
The combined result\cite{lep4f} shown in fig.\ref{zzcross} agrees well
with the SM predictions\cite{lepmc}. Combining all energies gave 
\begin{equation}
  {\sigma_{\rm ZZ} \over \sigma_{\rm ZZTO}} = 0.969 \pm 0.047(stat) \pm 0.028(syst),
  \label{avzz} 
\end{equation}
using ZZTO, one of the calculations. 
The theoretical uncertainty, estimated to be 2\%,
was higher than for W-pairs (because radiative corrections were not 
fully included) but sufficient given measurement errors. 
The main correlated systematic errors, arising from the background 
modelling, were smaller than the statistical ones. 

This measurement
constrained potential anomalous production from NP, as described
in sec.3.3 in the case of potential neutral TGCs. The reconstruction of
the Z-boson polar angle achieved by DELPHI in this context is shown 
in the lower plot of fig.\ref{delphintgc}. 

Reconstructing the Z-boson pairs at LEP-2 was also 
an important test for the Higgs searches, since
the analysed topologies were quasi-identical and cross-sections
similarly small. L3 measured the $\rm ZZ \rightarrow b {\bar b} X$ 
cross-section specially with this in mind\cite{l3zz}.

\subsection{Z$\gamma^*$ production}

Measurements of neutral boson pairs were also extended to include an
off-shell photon intead of a Z. The process can then be described as a 
``virtual radiative return to the Z'', with characteristic 
forward-peaked production and quasi mono-energetic $\gamma^*$
at the lower masses. 
Resulting topologies are distinctive and gave sizeable backgrounds in 
several searches for NP, which needed checking.
They also led to an original search for anomalous production via neutral TGCs, 
using a new parametrization extended to include off-shell terms (see sec.3.3).

\begin{figure}[!h]
  \vspace{8.5cm}
  \includegraphics{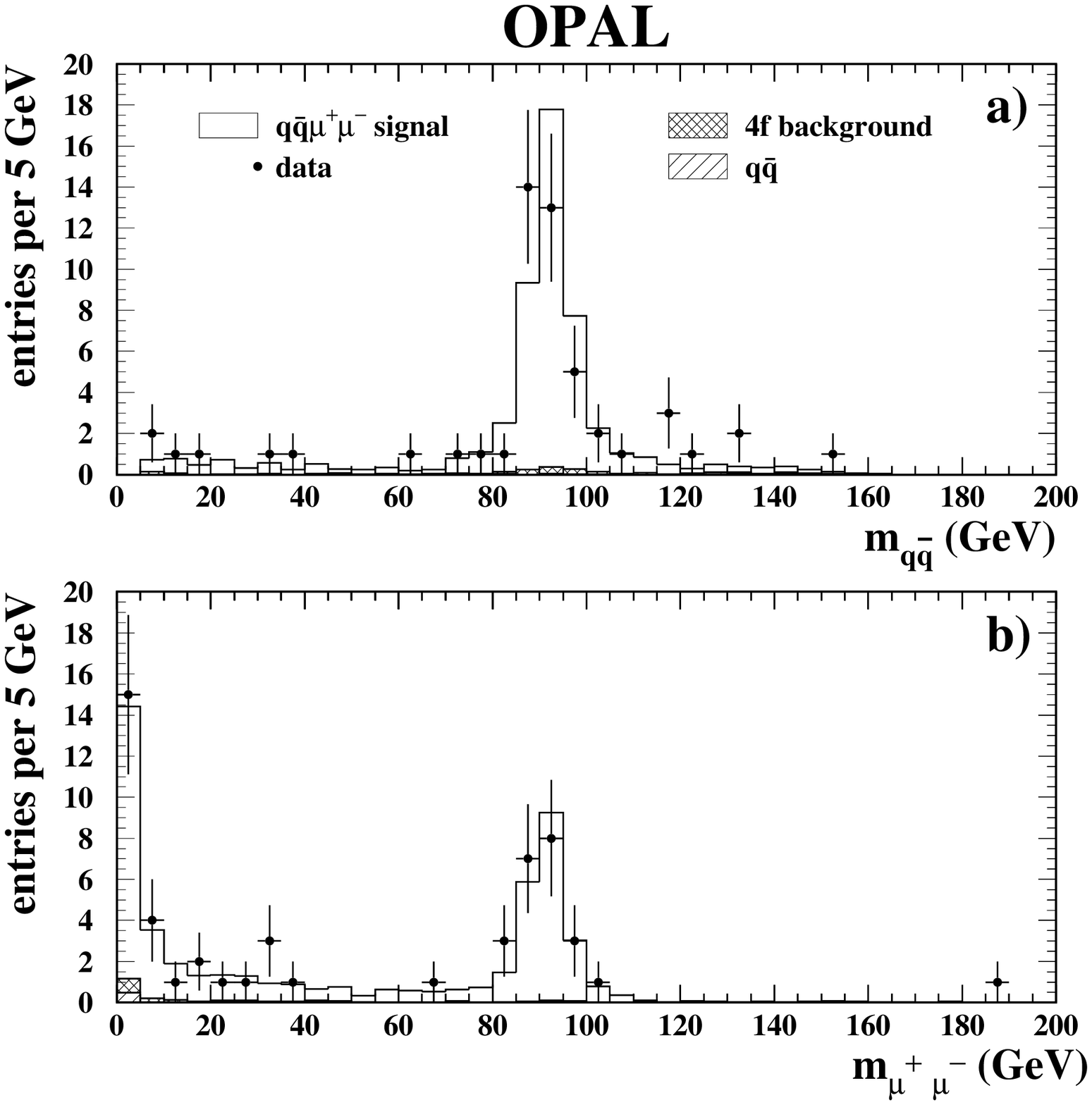}
  \includegraphics{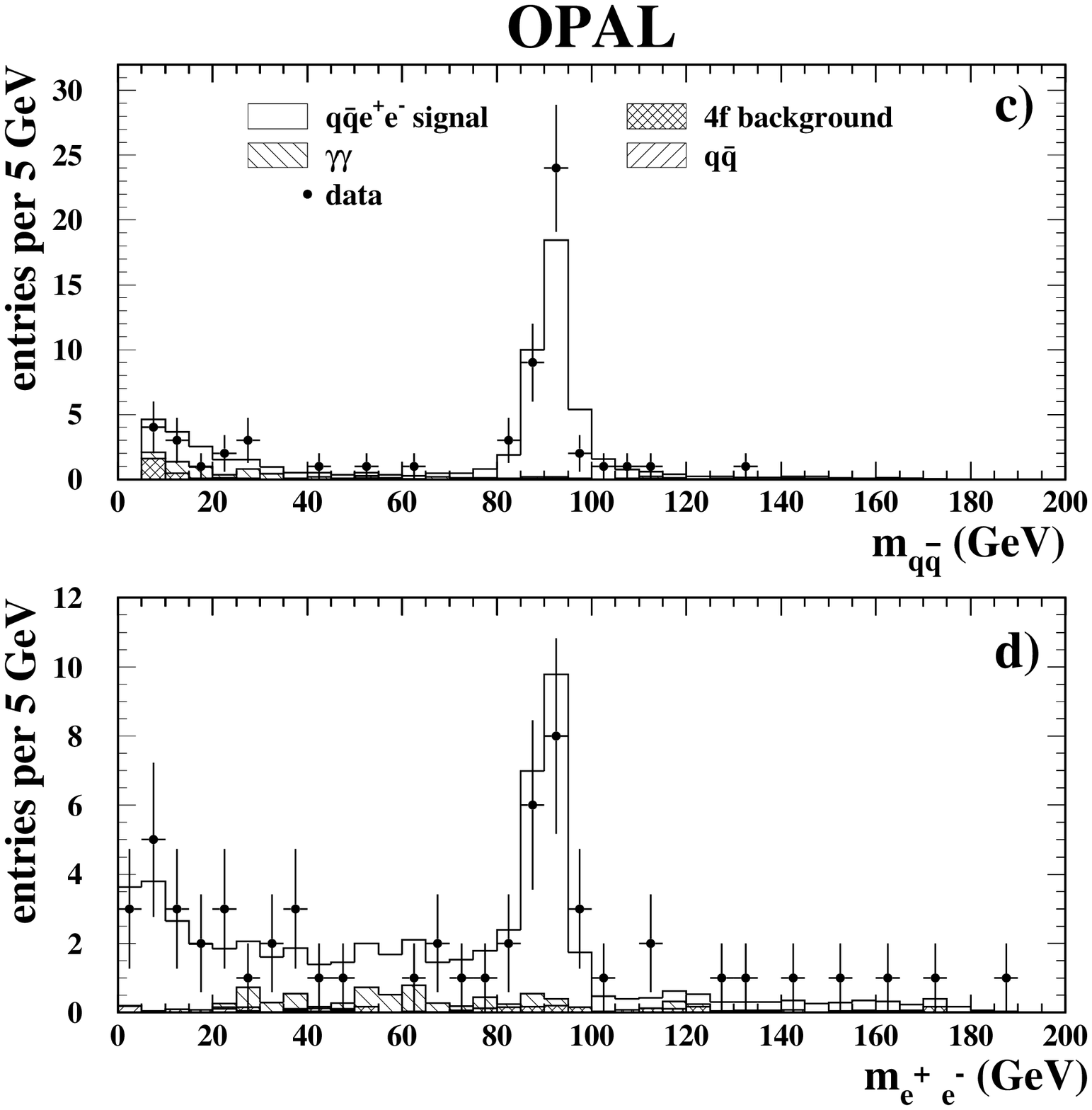}
  \caption{\it
    Di-fermion invariant masses obtained in the 
    $\rm Z\gamma^* \rightarrow l l q \bar{q}$ sub-channel
    after kinematic fitting with constraints from 
    four-momentum conservation.
    \label{zgopal} }
\end{figure}

DELPHI analysed the $\rm \mu \mu q \bar{q}$, 
$\rm e e q \bar{q}$, $\rm \nu \bar{\nu} q \bar{q}$, $\rm l l l l$
and $\rm q \bar{q} q \bar{q}$ final states and OPAL
the $\rm \mu \mu q \bar{q}$ and $\rm e e q \bar{q}$ ones\cite{opaldelphizg}.
Good agreement was found. Fig.\ref{zgopal} shows the di-quark and di-lepton 
mass spectra obtained by OPAL.
The suppression at low di-quark mass comes from the smaller 
leptonic Z-boson decay branching ratio. In final states with electrons, 
$t$-channel single-boson processes $\rm ee(Z/\gamma^*)$ with both
electron seen (see sec.2.5) enhanced the cross-section at high and 
low di-electron mass. 

Another interesting final state, 
$\rm \nu \bar{\nu} q \bar{q}$ has a 
mono-jet topology because the $\gamma^*$ mass 
distribution peaks at low values. 
The corresponding energy-averaged cross-section was
measured by DELPHI 
\begin{equation}
  \sigma_{\rm Z\gamma^* \rightarrow \nu \bar{\nu} q \bar{q}} = 
   0.129 \pm 0.035(stat) \pm 0.015(syst) {\rm ~pb}.
  \label{avzg} 
\end{equation}
The expected value was 0.088 pb. Uncertainties in the hadronization 
at low $\rm q \bar{q}$ mass and in the way to set $\alpha_{\rm em}$ given 
the different scales involved were shown to affect predictions
by up to 5\%\cite{wphact2}. 
Effects can be reduced by carefully treating the $t$-channel component 
in final states with electrons, and via phase space cuts. This should 
be good enough for the partial combination of three LEP experiments 
now planned.

\subsection{eeZ/$\gamma^*$ production}

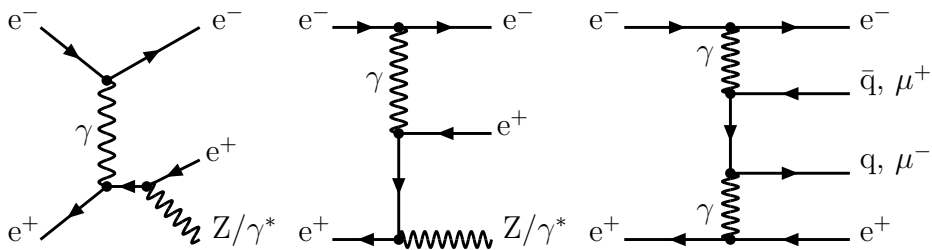
\begin{figure}[!h]
  \begin{center}
     \SetWidth{1.1}
     \begin{picture}(400,100)(0,100)
     \ArrowLine(30,180)(55,160)
     \Photon(55,160)(55,120){3}{6}
     \ArrowLine(55,120)(30,100)
     \ArrowLine(55,160)(90,180)
     \ArrowLine(70,120)(55,120)
     \ArrowLine(90,130)(70,120)
     \Photon(70,120)(90,100){3}{6}
     \Vertex(55,160){2}
     \Vertex(55,120){2}
     \Vertex(70,120){2}
     \Text(30,100)[rb]{e$^{+}$}
     \Text(30,180)[rb]{e$^{-}$}
     \Text(95,180)[lb]{e$^{-}$}
     \Text(105,130)[rb]{e$^{+}$}
     \Text(95,98)[lb]{Z/$\gamma^*$}
     \Text(50,140)[r]{$\gamma$}

     \ArrowLine(140,180)(165,180)
     \ArrowLine(165,180)(200,180)
     \Photon(165,180)(165,140){3}{8}
     \ArrowLine(165,140)(165,100)
     \ArrowLine(165,100)(140,100)
     \ArrowLine(200,140)(165,140)
     \Photon(165,100)(200,100){3}{8}
     \Vertex(165,140){2}
     \Vertex(165,180){2}
     \Vertex(165,100){2}
     \Text(140,100)[rb]{e$^{+}$}
     \Text(215,140)[rb]{e$^{+}$}
     \Text(140,180)[rb]{e$^{-}$}
     \Text(205,180)[lb]{e$^{-}$}
     \Text(205,98)[lb]{Z/$\gamma^*$}
     \Text(160,160)[r]{$\gamma$}

     \ArrowLine(250,180)(290,180)
     \ArrowLine(290,180)(335,180)
     \ArrowLine(335,100)(290,100)
     \ArrowLine(290,100)(250,100)
     \Photon(290,180)(290,155){3}{6}
     \Photon(290,100)(290,125){3}{6}
     \ArrowLine(290,155)(290,125)
     \ArrowLine(335,155)(290,155)
     \ArrowLine(290,125)(335,125)
     \Vertex(290,180){2}
     \Vertex(290,100){2}
     \Vertex(290,155){2}
     \Vertex(290,125){2}
     \Text(250,100)[rb]{e$^{+}$}
     \Text(250,180)[rb]{e$^{-}$}
     \Text(340,180)[lb]{e$^{-}$}
     \Text(340,100)[lb]{e$^{+}$}
     \Text(340,125)[lb]{q, $\mu^-$}
     \Text(340,155)[lb]{$\bar {\rm q}$, $\mu^+$}
     \Text(285,170)[r]{$\gamma$}
     \Text(285,110)[r]{$\gamma$}

  \end{picture}
  \vspace{1em}
  \caption{\it Examples of tree level Feynman graphs for single neutral boson production 
  (left and middle) and for the competing two-photon process (right).}
  \label{eezggraph}
  \end{center}
  \end{figure}

\begin{figure}[!h]
  \vspace{8.1cm}
  \includegraphics{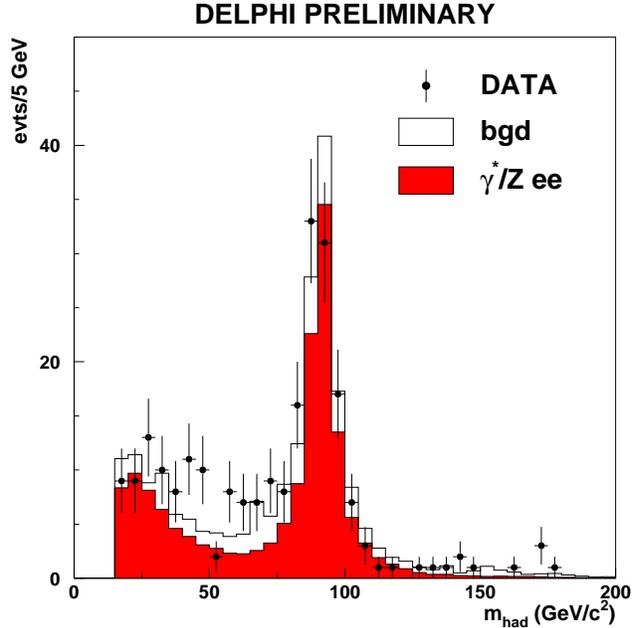}
  \caption{\it
    Hadronic invariant mass obtained in the reconstruction of the
    $\rm eeZ/\gamma^*$ process after kinematic fitting with constraints from 
    four-momentum conservation and assuming an electron lost 
    along the beam-line.
    \label{eezgmass} }
\end{figure}

Neutral bosons can be produced singly via the so-called
EW Compton scattering process $\rm e^+\gamma \rightarrow e^+\gamma^*/Z$,
where a quasi-real photon radiated from one of the beam electrons scatters off
the opposite one (see fig.\ref{eezggraph}).
The signature of such events is an electron in the detector with moderate energy
recoiling against the $\rm \gamma^*/Z$ system, with the other ``spectator'' electron
mostly lost in the beam-pipe. LEP collaborations measured the $\rm ee q {\bar q}$ 
and $\rm ee \mu \mu$ final states with one electron lost using 
data collected in 1997-2000\cite{confzee}. Competing ``single-tag'' contributions 
from two-photon processes were 
suppressed using correlations between the tag electron charge and direction. 
The reconstruction of
single-Z or $\gamma^*$ components is illustrated in fig.\ref{eezgmass},
where the hadronic mass spectrum from DELPHI is shown. The excess of data 
below the Z-boson mass is assumed to come from biases in the 
two-photon background. 
A combination of cross-sections using the ALEPH, DELPHI and L3 
results\cite{lep4f} was performed in the high mass single-Z region, giving good
agreement with expectations.  The signal was defined
as the cross-section 
from all graphs in the kinematic region: $m_{\rm f {\bar f}} >$ 60~GeV/$\rm c^2$ 
($\rm f =q,\mu$), $12^\circ < \theta_{\rm e^-} < 120^\circ$, 
$\theta_{\rm e^+} < 12^\circ$, $E_{\rm e^-} > 3$~GeV/$\rm c^2$ 
(with implicit charge conjugation). Averaging over energies gave
\begin{equation}
  {\sigma_{\rm Zee} \over \sigma_{\rm WPHACT}} = 0.951 \pm 0.068(stat) \pm 0.048(syst),
  \label{avzee} 
\end{equation}
using WPHACT, one of the calculations. 
Uncertainties in the way to set $\alpha_{\rm em}$ and from treating initial state 
radiation given the different scales involved affected the predictions 
at the 5\% level\cite{wphact2}. This matched the experimental errors, though
some differential effects may need to be included.

\subsection{$\rm e \nu_e W$ production}

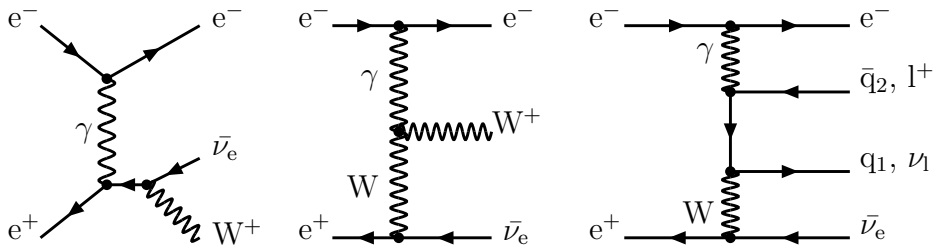
\begin{figure}[!h]
  \begin{center}
     \SetWidth{1.1}
     \begin{picture}(400,100)(0,100)
     \ArrowLine(30,180)(55,160)
     \Photon(55,160)(55,120){3}{6}
     \ArrowLine(55,120)(30,100)
     \ArrowLine(55,160)(90,180)
     \ArrowLine(70,120)(55,120)
     \ArrowLine(90,130)(70,120)
     \Photon(70,120)(90,100){3}{6}
     \Vertex(55,160){2}
     \Vertex(55,120){2}
     \Vertex(70,120){2}
     \Text(30,100)[rb]{e$^{+}$}
     \Text(30,180)[rb]{e$^{-}$}
     \Text(95,180)[lb]{e$^{-}$}
     \Text(105,130)[rb]{$\bar {\nu_{\rm e}}$}
     \Text(95,98)[lb]{W$^+$}
     \Text(50,140)[r]{$\gamma$}

     \ArrowLine(140,180)(165,180)
     \ArrowLine(165,180)(200,180)
     \Photon(165,180)(165,140){3}{8}
     \Photon(165,140)(165,100){3}{8}
     \ArrowLine(165,100)(140,100)
     \Photon(200,140)(165,140){3}{8}
     \ArrowLine(200,100)(165,100)
     \Vertex(165,140){2}
     \Vertex(165,180){2}
     \Vertex(165,100){2}
     \Text(140,100)[rb]{e$^{+}$}
     \Text(220,140)[rb]{W$^{+}$}
     \Text(140,180)[rb]{e$^{-}$}
     \Text(205,180)[lb]{e$^{-}$}
     \Text(205,98)[lb]{$\bar {\nu_{\rm e}}$}
     \Text(158,160)[r]{$\gamma$}
     \Text(158,120)[r]{W}

     \ArrowLine(250,180)(290,180)
     \ArrowLine(290,180)(335,180)
     \ArrowLine(335,100)(290,100)
     \ArrowLine(290,100)(250,100)
     \Photon(290,180)(290,155){3}{6}
     \Photon(290,100)(290,125){3}{6}
     \ArrowLine(290,155)(290,125)
     \ArrowLine(335,155)(290,155)
     \ArrowLine(290,125)(335,125)
     \Vertex(290,180){2}
     \Vertex(290,100){2}
     \Vertex(290,155){2}
     \Vertex(290,125){2}
     \Text(250,100)[rb]{e$^{+}$}
     \Text(250,180)[rb]{e$^{-}$}
     \Text(340,180)[lb]{e$^{-}$}
     \Text(340,100)[lb]{$\bar {\nu_{\rm e}}$}
     \Text(340,125)[lb]{${\rm q_1}$, ${\nu_{\rm l}}$}
     \Text(340,155)[lb]{$\bar {\rm q}_2$, l$^+$}
     \Text(285,170)[r]{$\gamma$}
     \Text(285,110)[r]{W}

  \end{picture}
  \vspace{1em}
  \caption{\it Examples of tree level Feynman graphs for single W-boson production
               (left and middle) and for the multiperipheral contribution (right).} 
  \label{wenugraph}
  \end{center}
  \end{figure}

W-bosons can also be produced singly via EW Compton scattering processes
$\rm e^+\gamma \rightarrow {\bar \nu_e} W^+$ as depicted in fig.\ref{wenugraph}.
The middle graph involves charged TGCs which can be probed measuring this process 
(see sec.3.2). As for single Z-boson production, the spectator electron is mostly lost in
the beam-pipe. 

\begin{figure}[!h]
  \vspace{5.2cm}
  \includegraphics{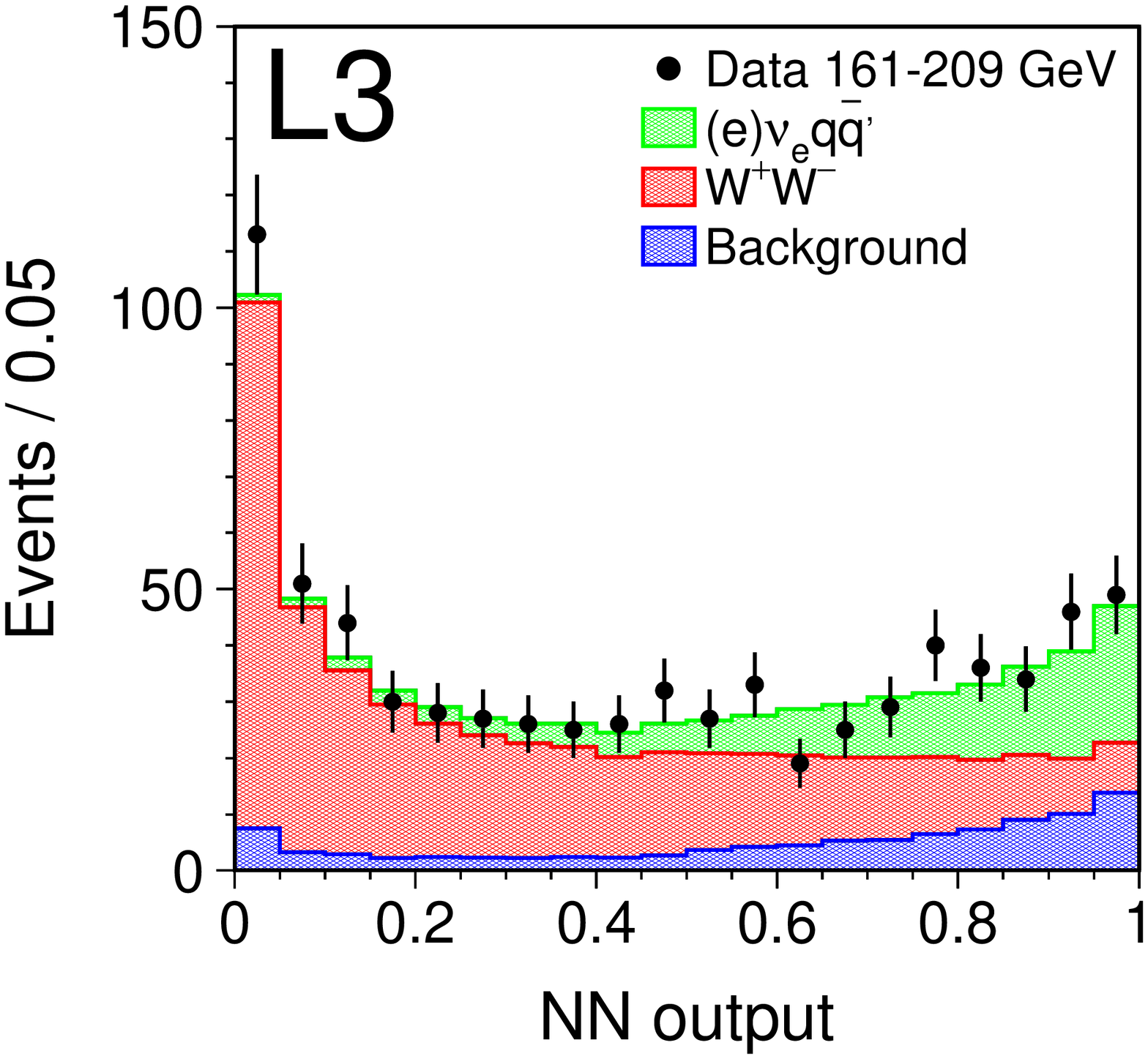}
  \includegraphics{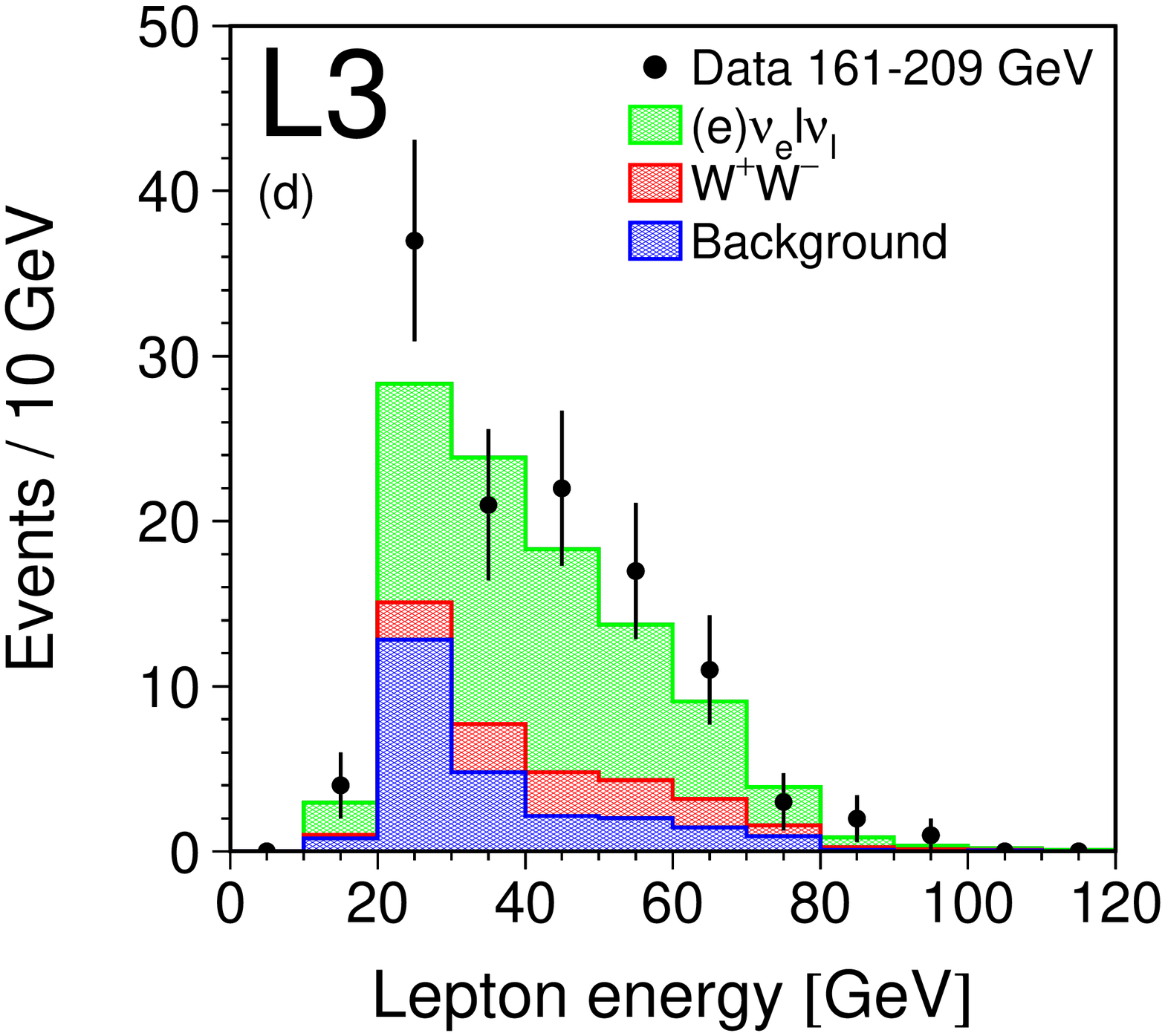}
  \caption{\it
    Output from the neural network used to isolate the single W-boson signal 
    in the hadronic sub-channel (left) and combined reconstructed lepton energy
    spectrum in the fully leptonic sub-channel (right).
    \label{l3wenu} }
\end{figure}

LEP collaborations measured all the possible final states
$\rm e \nu_e q \bar{q}$, $\rm e \nu_e l \nu_l $ ($\rm l=e,\mu,\tau$)
with the electron lost using data collected in 1997-2000\cite{confwenu}.
The main signature was the large missing energy and either a pair 
of acoplanar jets with mass close to $m_{\rm W}$ or a single 
energetic lepton. 
The reconstruction achieved by L3 is illustrated in 
fig.\ref{l3wenu}. The signal was defined as the complete $t$-channel
sub-set of $4f$ graphs within kinematic cuts specified to reduce theoretically 
poorly known multiperipheral contributions from graphs such as the last one 
in fig.\ref{wenugraph}: 
$\rm e \nu_e q \bar{q}$: $m_{\rm q\bar q}>45$ GeV/$\rm c^2$;
$\rm e \nu_e l \nu_l $ ($\rm l=\mu,\tau$): $E_{\rm l}>20$ GeV; 
$\rm e \nu_e e \nu_e $: $|\cos \theta_{\rm e^+}|<0.95$ and $E_{\rm e}^+>20$ GeV.
All results\cite{lep4f} agreed with SM
expectations\cite{lepmc}. 
Combining the cross-sections measured by all experiments at all energies gave 

\begin{equation}
  {\sigma_{\rm e \nu_e W} \over \sigma_{\rm GRACE}} = 0.949 \pm 0.067(stat) \pm 0.040(syst),
  \label{avwenu} 
\end{equation}
using GRACE, one of the calculations. 
Uncertainties similar to those described in sec.2.5 for single Z-bosons 
affected single W-bosons. Even though for the (gauge-invariant) 
$t$-channel graphs defining the single W-boson signal, complete fermionic 
one-loop corrections exist\cite{fermionloop}, 
allowing the correct scale for $\alpha_{\rm em}$ to be set, a 5\% error was 
estimated\cite{lepmc}, also here matching the experimental precision.

\subsection{Summary}

The comprehensive $4f$ measurement program conducted at LEP has been a success. It
provided a large set of original results and established experimentally the SM
environment where NP searches were carried out. The experimental precision achieved 
was matched by the accuracy of theory, as recalled in tab.\ref{xstab}, in
some cases after substantial work by the theoretical community\cite{lepmc}. 

In the case
of the W-pairs the non-Abelian gauge group structure of the SM
was clearly confirmed. The accuracy obtained, close to 0.5\%, even allowed the
SM calculation at loop level to be probed. In the case of the Z-pairs, 
a valuable experimental cross-check of the Higgs search at LEP-2 was made.
In the case of the single-resonant boson processes SM predictions were tested in
several yet unexplored regions. 

Not all topics covered by the $4f$ sub-group of the LEP EW working group 
or by individual experiments could be reviewed in this report, for instance the
measurements of the production polar angle, decay branching ratios, polarisation 
and spin correlations in the W-boson pair production process, and of 
the Z$\gamma \gamma$ cross-section\footnote{A first partial combination of WW$\gamma$
cross-section measurements is shown in fig.\ref{lepqgc} in the context of the search
for QGCs described in sec.3.4.}.

\begin{table}[!h]
  \centering
  \caption{ \it Overview of experimental results and estimated theory errors
                for the production processes $e^+e^- \rightarrow WW$, $ZZ$, 
                $eeZ/\gamma^*$ and $e\nu W$.
                The predictions in the quoted ratios were obtained with the 
                event generators YFSZZ, ZZTO, WPHACT and GRACE, 
                respectively\cite{lepmc}.
    }
  \vskip 0.1 in
  \begin{tabular}{|l|c|c|} \hline
    physical process & measurement / prediction & theoretical precision             \\
    \hline
    \hline
    $\rm e^+e^- \rightarrow WW$ & $ 0.997 \pm 0.007(stat) \pm 0.009(syst) $ & 0.005 \\
    $\rm e^+e^- \rightarrow ZZ$ & $ 0.969 \pm 0.047(stat) \pm 0.028(syst) $ & 0.02 \\
    $\rm e^+e^- \rightarrow eeZ/\gamma^*$ 
                                & $ 0.951 \pm 0.068(stat) \pm 0.048(syst) $ & 0.05 \\
    $\rm e^+e^- \rightarrow e\nu_eW$ 
                                & $ 0.949 \pm 0.067(stat) \pm 0.040(syst) $ & 0.05 \\
    \hline
  \end{tabular}
  \label{xstab}
\end{table}
\section{Gauge boson self-couplings}
\subsection{Overview}

Couplings between the SM gauge bosons were measured at LEP by analysing 
$4f$ (and other) final states. Deviations from tree level values are 
predicted in NP scenarios and arise also from radiative corrections. If large 
enough to be measured, such deviations can help to probe NP at energy scales 
beyond the kinematic range of direct searches for new particles. 

Anomalous 
self-couplings were searched for at three and four boson vertices, involving 
both charged and neutral gauge bosons. Charged TGCs exist in the SM due to 
the non-Abelian gauge group structure (see sec.2.2). On the contrary neutral 
TGCs vanish at tree level. The SM predicts QGCs, but their size is very 
small. In the two latter cases one can only hope to detect anomalous contributions.
\subsection{Charged triple gauge couplings}

Parametrizations for the tree level VWW vertices (V=Z,$\gamma$) in the right and
middle graphs of fig.\ref{cc03graph} and \ref{wenugraph} involve in their
most general form a Lorentz invariant Lagrangian with fourteen independent complex 
couplings\cite{hagiwara}. Restricting the search to models with symmetries as in 
the SM (C, P, U(1)$_{em}$ and $\rm SU(2)_L \times U(1)_Y$) the number of independent 
couplings reduces to three: $\rm g^Z_1$, $\kappa_{\gamma}$ and $\lambda_{\gamma}$. 
In the SM the two first couplings are equal to 1 and the latter vanishes. They can 
be related to the weak charge, magnetic dipole and electric quadrupole moments of 
the W-boson. 

\begin{figure}[!t]
  \vspace{4.8cm}
  \includegraphics{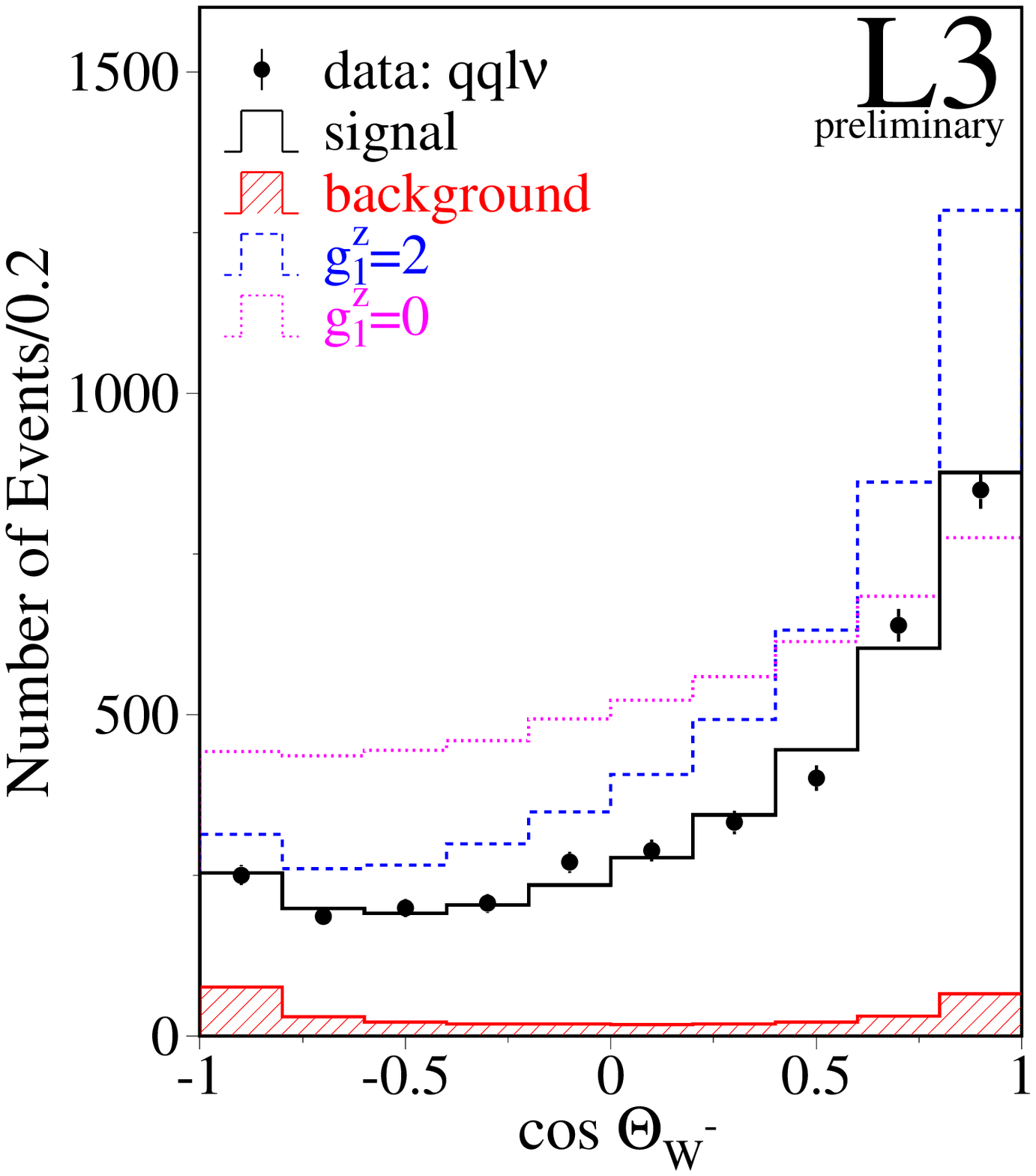}
  \includegraphics{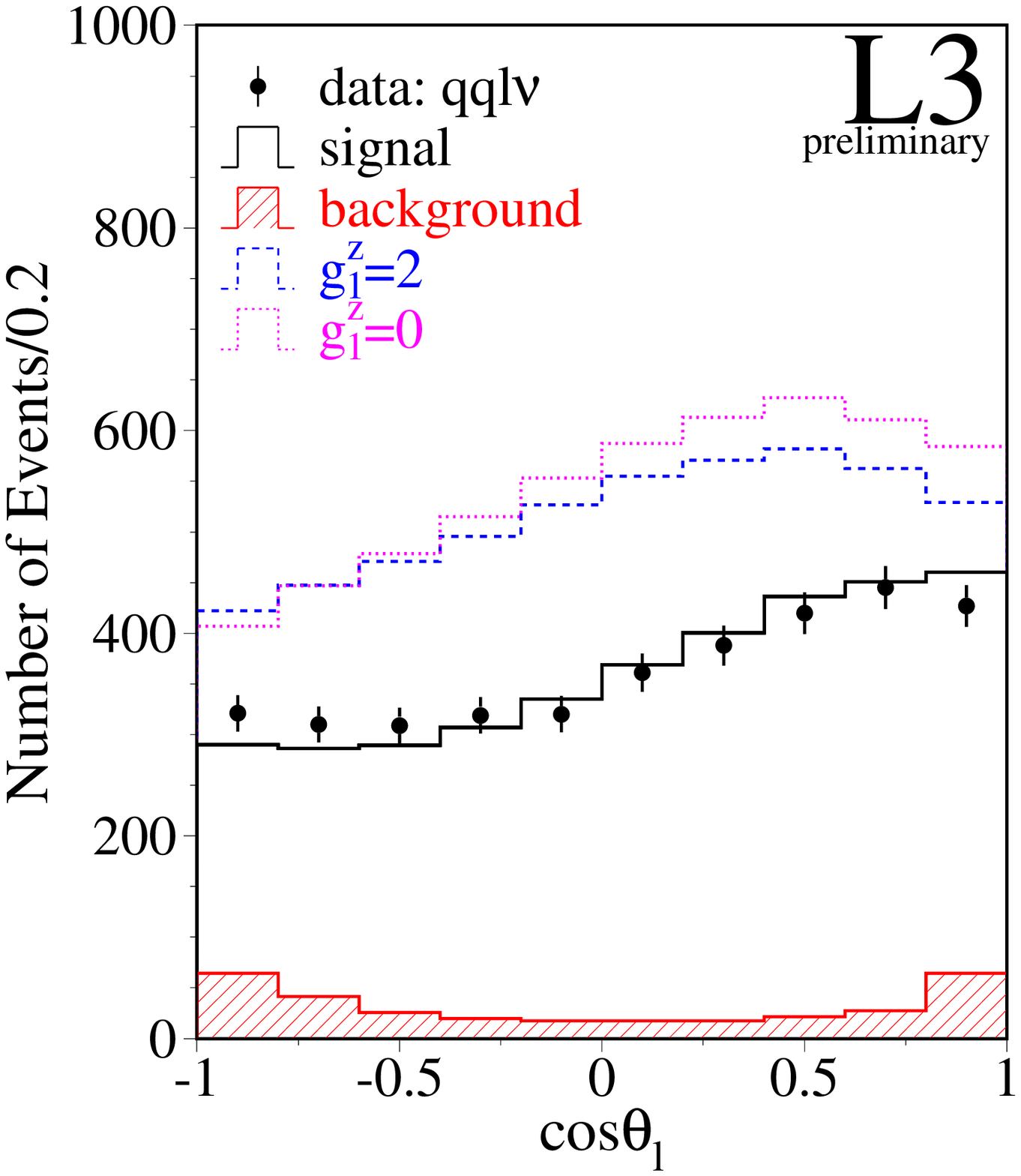}
  \includegraphics{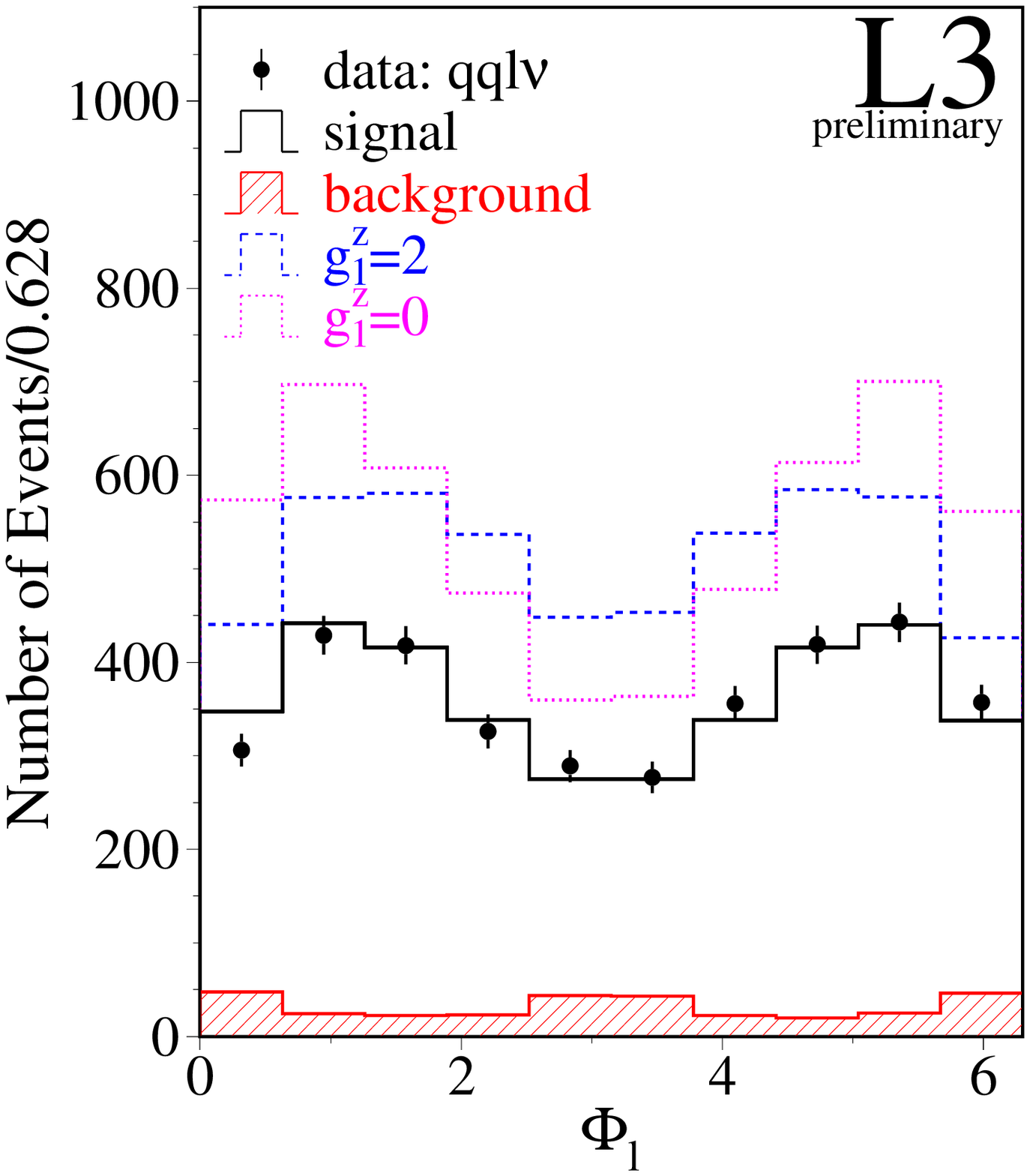}
  \caption{\it
    Measured production polar angles (left), decay polar angles (middle) and 
    decay azimuthal angles (right) of the W-bosons in the sub-channel
    $\rm WW \rightarrow \rm l \nu q \bar{q}$. The sensitivity to deviations from the
    SM prediction $\rm g^Z_1=1$ is indicated. 
    \label{l3ctgc} }
\end{figure}

Deviations from the SM values can be probed in a complementary way via 
effects induced on the single and pair production processes (see sec.2.6 and 2.2), 
the former being sensitive mainly to $\kappa_{\gamma}$ and the latter mainly to 
$\rm g^Z_1$ and $\lambda_{\gamma}$\footnote{
The $\nu \nu \gamma$ final state also has some sensitivity to $\kappa_{\gamma}$ and 
$\lambda_{\gamma}$ through the WW fusion process\cite{opalgnn}.}. 
Results obtained by the four LEP collaborations using the data collected in 
1996-2000 are reported in\cite{confctgc}. To maximize the sensitivity, the 
cross-sections, boson production polar angles, decay polar and azimuthal 
angles and average polarisation were exploited, using several adapted
combination methods, based for example on ``optimal observables''\cite{optobs}. 
The reconstruction performed by L3 for the boson production and 
decay angles in semi-leptonic W-boson pairs is illustrated in fig.\ref{l3ctgc}.

\begin{figure}[!h]
  \vspace{11.4cm}
  \includegraphics{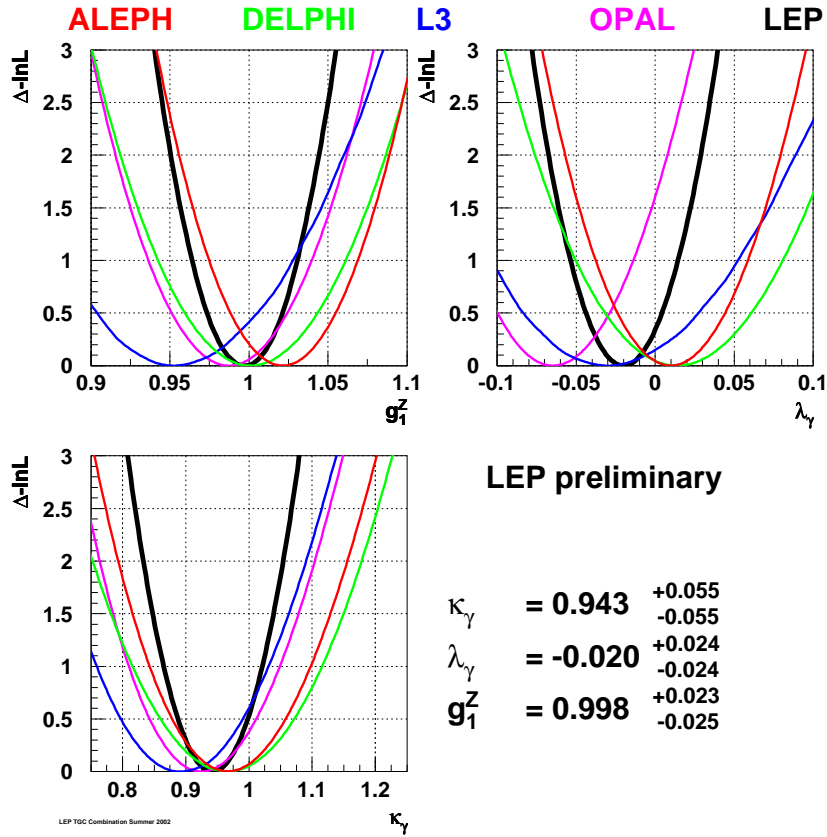}
  \caption{\it
    Measurement of the charged TGC parameters $\rm g^Z_1$, $\kappa_{\gamma}$ and 
    $\lambda_{\gamma}$.  
    \label{lepctgc} }
\end{figure}

The measurement of charged TGCs obtained by combining all LEP results is shown 
in fig.\ref{lepctgc}\cite{lep4f} with the negative log likelihoods 
provided by each experiment for each coupling, and their sums. Each coupling was 
minimized independently with the two others kept at their SM values\footnote{
Two and three-parameter analyses resulted in weak correlations.}. 
Good agreement with the tree-level SM expectations was found within errors of 2-5\%. 

The main source of correlated systematic uncertainty was of theoretical nature, from 
the recent inclusion of $O(\alpha)$ EW radiative corrections in the W-pair production 
process (see sec.2.2). This source of error was estimated conservatively as the full 
difference between results using Monte Carlo samples with and without 
these new corrections, yielding negative shifts of -0.015 for $\rm g^Z_1$
and $\lambda_{\gamma}$ and -0.04 for $\kappa_{\gamma}$.
Although 
these shifts are sizeable, it can be seen from their signs and from the values 
found for the couplings that the agreement with the SM is as good with or 
without them. This is different from the cross-section 
measurement, where the improvement from the new corrections was clear. This feature
is perhaps a little surprising, since more information is used to extract the charged TGCs,
like angles and the single W-boson channel.

Since these shifts had similar sizes as the statistical errors and were fully 
correlated between energies and experiments, a careful treatment was implemented in the 
combination, expressing the likelihoods as functions of each coupling and of additional 
free parameters, to represent each error weighted by its sensitivity in each experiment, 
and by then performing a simultaneous minimization\cite{lep4f}. 

\subsection{Neutral triple gauge couplings}

There are no self-couplings involving exclusively neutral gauge bosons at tree level in the SM.

For NP satisfying Lorentz invariance and preserving the U(1)$_{em}$ and Bose symmetries for 
identical particles, the most general parametrization for the ZZZ, ZZ$\gamma$ and Z$\gamma \gamma$ 
vertices has twelve independent parameters, six of which are CP-conserving: $h_{3,4}^{\rm Z/\gamma}$ 
and $f_5^{\rm Z/\gamma}$, while the other six are CP-violating: $h_{1,2}^{\rm Z/\gamma}$ and 
$f_4^{\rm Z/\gamma}$\cite{hagiwara}. The $h$ and $f$ terms describe the VZ$\gamma$ and VZZ vertices, 
respectively, where $\rm V = Z,\gamma$ is off-shell but the two other bosons are on-shell. 

\begin{figure}[!h]
  \vspace{14.0cm}
  \includegraphics{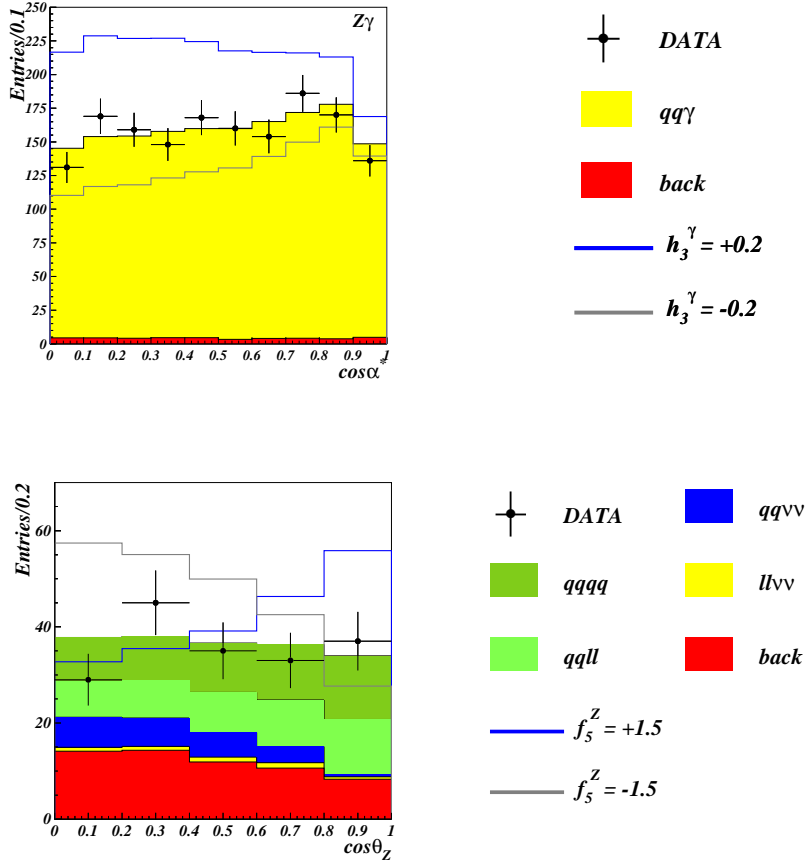}
  \caption{\it
    Measured decay angle of the Z in its rest frame in the  
    $\rm e^+e^- \rightarrow Z \gamma \rightarrow q \bar{q} \gamma$
    process (upper plot).
    Measured production angle in Z-boson pairs for the main final states studied by 
    DELPHI\cite{confzz} (lower plot).
    The sensitivity to the presence of CP-conserving $h_3^{\gamma}$ and $f_5^{\rm Z}$ 
    couplings, respectively at the $\gamma$Z$\gamma$ and ZZZ vertices, is indicated.
    \label{delphintgc} }
\end{figure}

These two classes of neutral TGCs can 
be probed via effects induced on the production processes $\rm e^+e^- \rightarrow Z \gamma$ and ZZ,  
respectively (see sec.2.3 for the measurement of the latter).
Results obtained by the four LEP collaborations using the data collected in 
1996-2000 are reported in\cite{confntgc}. The cross-sections and boson production angles were 
exploited, as well as, for the $\rm e^+e^- \rightarrow Z \gamma \rightarrow q \bar{q} \gamma$
process the angle between the photon and nearest jet and, for the 
$\rm e^+e^- \rightarrow Z \gamma \rightarrow \nu \bar{\nu} \gamma$ process the photon energy.
The reconstruction performed by DELPHI for the two angles just described is illustrated
in fig.\ref{delphintgc}. 

The constraints on neutral TGCs obtained combining all LEP results are shown 
in tab.\ref{ntgctab}\cite{lep4f}. The precision was dominated by statistical errors.
There was lower sensitivity to $f$ than to $h$ terms because of the less favourable 
kinematics at the threshold for Z-boson pair production. 

\begin{table}[!h]
  \centering
  \caption{ \it Combined 95 \% confidence level probability intervals for the neutral TGCs probed
    by measuring the $\rm e^+e^- \rightarrow Z \gamma$ and ZZ processes. The constraints shown
    on $h$ and $f$ terms result from the single and two-parameter analyses, respectively. 
    Some correlation between $h$ terms was observed in a two-parameter analysis done combining 
    results from three experiments\cite{lep4f}. 
    The two latter bosons at each listed vertex correspond to the final state produced and 
    are on-shell.
    }
  \vskip 0.1 in
  \begin{tabular}{|l|c|c|c|} \hline
    vertex & CP & parameter & 95 \% CL limits             \\
    \hline
    \hline
    $\rm \gamma Z\gamma$ &  odd & $h_1^{\rm \gamma}$   & [-0.056, +0.055] \\
    \hline
    $\rm \gamma Z\gamma$ &  odd & $h_2^{\rm \gamma}$   & [-0.045, +0.025] \\
    \hline
    $\rm \gamma Z\gamma$ & even & $h_3^{\rm \gamma}$   & [-0.049, +0.008] \\
    \hline
    $\rm \gamma Z\gamma$ & even & $h_4^{\rm \gamma}$   & [-0.002, +0.034] \\
    \hline
    $\rm Z Z\gamma$      &  odd & $h_1^{\rm Z}$   & [-0.13, +0.13] \\
    \hline
    $\rm Z Z\gamma$      &  odd & $h_2^{\rm Z}$   & [-0.078, +0.071] \\
    \hline
    $\rm Z Z\gamma$      & even & $h_3^{\rm Z}$   & [-0.20, +0.07] \\
    \hline
    $\rm Z Z\gamma$      & even & $h_4^{\rm Z}$   & [-0.05, +0.12] \\
    \hline
    \hline
    $\rm \gamma ZZ$ &  odd  & $f_4^{\rm \gamma}$   & [-0.17, +0.19] \\
    $\rm       ZZZ$ &  odd  & $f_4^{\rm Z}$        & [-0.30, +0.28] \\
    \hline
    $\rm \gamma ZZ$ &  even & $f_5^{\rm \gamma}$   & [-0.34, +0.38] \\
    $\rm       ZZZ$ &  even & $f_5^{\rm Z}$        & [-0.36, +0.38] \\
    \hline
  \end{tabular}
  \label{ntgctab}
\end{table}

The analysis done 
treated $h$ and $f$ terms separately, though it has recently been shown that using the
$\rm SU(2)_L \times U(1)_Y$ symmetry as in the charged case relates the VZ$\gamma$ 
and VZZ vertices for some classes of operators and can lead to 
fewer independent couplings\cite{alcaraz}. 
Although some generality may be lost, this 
will now be done as well. It is expected that the more precise
$\rm e^+e^- \rightarrow Z \gamma$ measurements will dominate the constraints from such a 
combined analysis. 

Another recent theoretical development has enabled a generalization of the description by
including off-shell bosons\cite{gounaris}.
It was argued that the resulting effects cannot be ignored in detailed experiments, 
especially if data measured outside the strictly on-shell regions of the Z-pair and 
Z$\gamma$ processes are also used. A first study in this direction has been presented
by DELPHI, based on the Z$\gamma^*$ production measurement (see sec.2.4)\cite{confntgc}.

\subsection{Quartic gauge couplings}

The couplings predicted in the SM at the WWWW, WWZZ, WW$\gamma\gamma$ and WWZ$\gamma$ 
vertices are below LEP-2 sensitivities. The searches performed probed potential 
anomalous contributions arising from NP, concentrating on operators which do not 
simultaneously cause anomalous TGCs. It has been argued that such operators, often 
referred to as ``genuine'' QGCs, can be related more directly to the scalar sector 
of the theory\cite{godfrey}. The parametrization used involves four CP-conserving terms at 
the WW$\gamma\gamma$ and ZZ$\gamma\gamma$ vertices, $a_{0,c}^{\rm W,Z}$, and a 
CP-violating one at the WWZ$\gamma$ vertex, $a_n$\cite{qgctheory}. By convention
these terms are usually normalised to $\Lambda^2$, the square of the energy scale
at which the NP responsible for them appears.

Experimentally these
terms were probed by measuring the three boson final state processes 
$\rm e^+e^- \rightarrow WW \gamma$ (for $a_{0,c}^{\rm W}$ and $a_n$)
and $\rm e^+e^- \rightarrow Z \gamma \gamma$ (for $a_{0,c}^{\rm Z}$), using both
the rates and energy spectra of the photons\cite{3bosonqgc}. The presence of $a_{0,c}^{\rm W,Z}$ 
terms was also studied through their influence on the $\rm e^+e^- \rightarrow \nu\nu\gamma\gamma$ 
process, via a contributing graph with WW fusion and another one involving neutrino pair 
production mediated by a Z boson which radiates two photons\cite{opalnngg}. Both the rate 
and the photon pair recoil mass spectrum were exploited. 

\begin{figure}[!h]
  \vspace{9.2cm}
  \includegraphics{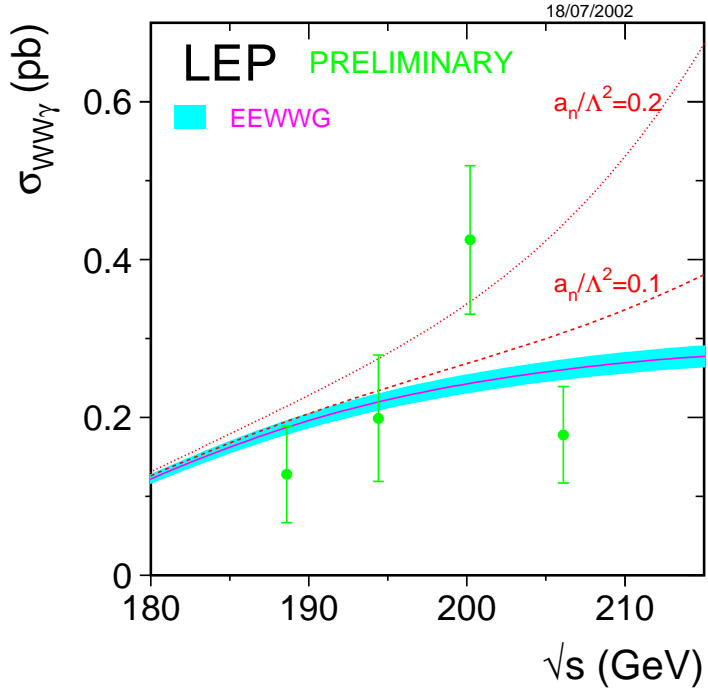}
  \caption{\it
    Combination of WW$\gamma$ cross-section measurements by DELPHI and L3 compared with 
    the SM prediction. The sensitivity to the presence of a CP-violating 
    $a_n$ coupling at the WWZ$\gamma$ vertex is indicated by the predicted 
    distribution for two values of $a_n/\Lambda^2$ (in units of GeV$^{-2}$).
    \label{lepqgc} }
\end{figure}

Only the L3 Z$\gamma \gamma$ and OPAL $\nu\nu\gamma\gamma$ results were combined so far,
allowing to set the following 95\% CL limits on the ZZ$\gamma\gamma$ vertex: 
$-0.033 < a_{0}^{\rm Z} \times GeV^2 / \Lambda^2  < 0.046$ and 
$-0.009 < a_{c}^{\rm Z} \times GeV^2 / \Lambda^2  < 0.026$\cite{lep4f}.
A combination of the cross-sections obtained by DELPHI and L3 for the WW$\gamma$ final state 
was also performed\footnote{The signal was defined as the cross-section from all graphs
in the kinematic region: $|m_{\rm f {\bar f}} - m_{\rm W}| < 2 \Gamma_{\rm W}$,
$\cos\theta_{\gamma, \rm f} < 0.90$, $|\cos\theta_{\gamma}| < 0.95$ and $E_{\gamma} > 5$ GeV.} 
and is illustrated in fig.\ref{lepqgc}, together with the SM prediction of 
EEWWG\cite{eewwg} and the sensitivity to the CP-violating $a_n$ coupling. No combined 
constraints on 
$a_{0,c}^{\rm W}$ and $a_n$ QGCs were yet obtained from this limited cross-section 
information, but individual results have been published, for example the L3 one: 
$-0.015 < a_{0}^{\rm W} \times GeV^2 / \Lambda^2  < 0.015$ and 
$-0.048 < a_{c}^{\rm W} \times GeV^2 / \Lambda^2  < 0.026$ 
(using also the information from the $\nu\nu\gamma\gamma$ final state), 
and $-0.14 < a_n \times GeV^2 / \Lambda^2 < 0.13$. 

\subsection{Summary}

The charged TGCs were measured within a few 0.01 of their predicted values in the SM, 
confirming again the non-Abelian gauge group structure. NP giving anomalous contributions 
of the order of these errors could be excluded. The sensitivity was not enough to
probe SM loop effects, which are predicted to be at the 0.003 level. It was also barely enough
to sense effects from SUSY even in the most optimistic scenarios with sparticles not 
far above the kinematic limit. A Z$^\prime$ with a low mass would on the other hand have produced 
visible effects, but would also have been strongly felt in di-fermion cross-section 
measurements\cite{lep2yr96}.

Some improvement to the precision is still expected from on-going work to estimate 
the uncertainty on the $O(\alpha)$ corrections in a better way than just quoting 
their full effect, for instance by varying assumptions in the theoretical treatments
used.
This could help to understand
why there was some sensitivity to SM loop effects when measuring
the total cross-section (see sec.2.2) but not the charged TGCs 
(see sec.3.2).
Further ahead experiments at higher energies will improve the 
sensitivity. 
The TeVatron with 10 fb$^{-1}$ and the LHC with 
300 fb$^{-1}$ should for example pin down $\lambda_{\gamma}$ to about $\pm 0.005$ 
and $\pm 0.0003$, respectively. A good sensitivity to all TGCs is expected at a
future $e^+e^-$
linear collider. For instance TESLA should reach a precision of a few $\pm 0.0001$
after collecting 1500 fb$^{-1}$ at $\sqrt{s}$ = 800 GeV\cite{tesla}.

The neutral TGCs were found to be zero as expected, but within larger error ranging
from $\pm 0.05$ to $\pm 0.30$. In the case of perturbative NP, anomalous contributions 
to neutral TGCs are expected to be depressed by at least one more power of 
$m_{\rm W,Z}^2 / \Lambda^2$ compared to charged ones, because operators with higher 
dimension are involved. Neutral TGCs expected from several scenarios for NP have
for example been studied in \cite{ntgcgounaris}. The overall conclusion reached
was that potential effects were generally below experimental sensitivities, except in 
cases of new particles with masses just above the kinematic reach, or of NP which is
not perturbative.

The QGCs were also found to be close to zero as expected. Here the expected 
``natural'' size of couplings at LEP-2 is about 1 for NP with an energy scale
$\Lambda$ = 3 TeV, which is far below the experimental sensitivity\cite{belanger}.

Even though effects on gauge couplings from the NP scenarios that were investigated
turned out to be too small to be detected, the systematic search program carried out 
at LEP-2 was certainly justified to demonstrate the validity of the SM, and to show
that there is no evidence of new physics from totally unexpected sources. The analyses 
done are also a useful preparation for similar work at future high energy colliders.

Recent measurements and analyses of spin density matrix elements in the W-boson pair 
production process are not covered here.

\vspace{\fill}

\section{Conclusions and prospects}

The results on the measurements of $4f$ final states and gauge boson self-couplings
are an important part of the LEP legacy.
The present work is to complete the documentation promptly. 
while the main physicists 
involved are still available. 
All final experimental results and combinations are expected 
during 2003. At the time of this writing, only the measurements by DELPHI of Z-boson pair 
production, by L3 of single boson production and quartic couplings, and by OPAL of 
Z$\gamma^*$ production are 
considered
truly ``final''
in the sense of being described 
in a refereed CERN-EP preprint note or journal publication. 
It is important to ensure
a high quality and an appropriate level of detail in the descriptions of the analyses
and final results to facilitate future reading.

\section{Acknowledgements}

The beautiful experimental results presented in this report are one of the cherries on 
the LEP-2 cake: a long chain of prior tasks was required.
The accelerator physicists and operators achieved energies and luminosities
in excess of expectations. The experimental teams took quality data with ever rising 
efficiencies. Alignment and calibration procedures were carried out. Huge sets of 
Monte Carlo simulations were produced and checked. And more... All this detailed 
work and the people involved must be recognised. 

The work of the $4f$ and Gauge Couplings LEP working groups should also be 
acknowledged. Such common inter-collaboration groups provide a competitive 
and stimulating environment where the best experts from each experiments 
can share their knowledge, discuss approaches, compare and cross-scrutinize 
each others results and methods. They were an important ingredient which enhanced
the scientific quality and can serve as example for the future.

Personally, I would like to thank my colleagues in the DELPHI $4f$ team for 
collaborating over the years and for the good spirit, in particular 
Sandro Ballestrero, Marcia Begalli, Maurizio Bonesini, Guennadi Borisov, Roberto Contri, 
Niels Kjaer, 
Esther Ferrer, Enrico Graziani, Anna Lipniacka, Ernesto Migliore, Rosy Nikolaidou, 
Hywel Phillips, Maria Elena Pol, Jens Rehn, Robert Sekulin, Alessandra Tonazzo, 
Valerio Verzi, Ivo van Vulpen and Mariusz Witek.
Special thanks also to the successive DELPHI spokesmen
Daniel Treille, Wilbur Venus, Tiziano Camporesi and Jan Timmermans,
who supported our work and to Roberto Chierici, Ulrich Parzefall and Robert Sekulin, 
who helped me prepare this review. Jan, Robert, Roberto and Sandro also kindly 
proof-read this text.

\vspace{\fill}

\end{document}